\documentclass[twocolumn]{aastex631}

\newcommand{\bs}{\boldsymbol}

\usepackage{amsmath}    
\usepackage{amssymb}    
\usepackage{xcolor}

\begin{document}


\title{Observational constraints of radial migration in the Galactic disc driven by the slowing bar}

\author[0009-0005-6898-0927]{HanYuan Zhang}
\affiliation{Institute of Astronomy, University of Cambridge, Madingley Road, Cambridge CB3 0HA, UK}

\author[0000-0002-0038-9584]{Vasily Belokurov}
\affiliation{Institute of Astronomy, University of Cambridge, Madingley Road, Cambridge CB3 0HA, UK}

\author[0000-0002-5981-7360]{N. Wyn Evans}
\affiliation{Institute of Astronomy, University of Cambridge, Madingley Road, Cambridge CB3 0HA, UK}

\author[0000-0003-4593-6788]{Jason L. Sanders}
\affiliation{Department of Physics and Astronomy, University College London, London WC1E 6BT, UK}

\author[0000-0003-4769-3273]{Yuxi(Lucy) Lu}
\affiliation{Department of Astronomy, The Ohio State University, Columbus, 140 W 18th Ave, OH 43210, USA}
\affiliation{Center for Cosmology and Astroparticle Physics (CCAPP), The Ohio State University, 191 W. Woodruff Ave., Columbus, OH 43210, USA}

\author[0000-0001-6655-854X]{Chengye Cao}
\affiliation{Department of Astronomy, School of Physics and Astronomy, 800 Dongchuan Road, Shanghai Jiao Tong University, Shanghai 200240, People's Republic of China \\
}

\author[0000-0002-5629-8876]{GyuChul Myeong}
\affiliation{Institute of Astronomy, University of Cambridge, Madingley Road, Cambridge CB3 0HA, UK}

\author[0000-0003-0807-5261]{Adam M. Dillamore}
\affiliation{Department of Physics and Astronomy, University College London, London WC1E 6BT, UK}

\author[0000-0001-8411-1012]{Sarah G. Kane}
\affiliation{Institute of Astronomy, University of Cambridge, Madingley Road, Cambridge CB3 0HA, UK}

\author[0000-0001-5017-7021]{Zhao-Yu Li}
\affiliation{Department of Astronomy, School of Physics and Astronomy, 800 Dongchuan Road, Shanghai Jiao Tong University, Shanghai 200240, People's Republic of China \\
}
\affiliation{National Key Laboratory of Dark Matter Physics, School of Physics and Astronomy, Shanghai Jiao Tong University, Shanghai 200240, People's Republic of China\\}

\correspondingauthor{HanYuan Zhang}
\email{hz420@cam.ac.uk}



\begin{abstract}
Radial migration is an important dynamical effect that has reshaped the Galactic disc, but its origin has yet to be elucidated. In this work, we present evidence that resonant dragging by the corotation of a decelerating bar could be the main driver of radial migration in the Milky Way disc. Using a test particle simulation, we demonstrate this scenario
explains the two distinct age-metallicity sequences observed in the solar vicinity: the plateauing upper sequence is interpreted as stars dragged outwards by the expanding corotation of the decelerating bar and the steeper lower sequence as stars formed locally around the solar circle. 
The upper migrated sequence dominates at guiding radii around the current corotation radius of the bar, $R\sim7\,\mathrm{kpc}$, but rapidly dies away beyond this where the mechanism cannot operate.
This behaviour naturally explains the radial dependence of the $\mathrm{[\alpha/Fe]}$-bimodality, in particular the truncation of the high-$\mathrm{[\alpha/Fe]}$ disc beyond the solar circle. Under our proposed radial migration scenario, we constrain the Milky Way bar's pattern speed evolution using the age-metallicity distribution of stars currently trapped at corotation. We find the bar likely formed with an initial pattern speed of $60-100$~km s$^{-1}$ kpc$^{-1}$ and began decelerating $6-8$~Gyr ago at a rate $-\dot{\Omega}/\Omega^2\sim0.0025-0.0040$ (where the quoted ranges include systematic uncertainties).

\end{abstract}



\section{Introduction} \label{sec:intro}

Stars in the galactic disc can be scattered from their presumed near-circular birth orbits to their present-day orbits by secular processes or external perturbations. This change in stellar orbits can be decomposed into two different components, the drifting of the angular momentum ($\Delta L_z$, dubbed ``radial migration", \citealt{Sellwood_Binney2002}, or ``churning", \citealt{SchonrichBinney2009}), and the amplification in the radial oscillation or the radial action ($\Delta J_R$, dubbed ``radial heating" or ``blurring").

Many studies have shown that radial migration in the Milky Way disc is strong using age and metallicity of various kinds of tracers \citep[e.g.][]{SchonrichBinney2009, SandersBinney2015,Frankel2018, BeraldoeSilva2021, Zhang2021, Lian2022_rm, Haywood2024}. The age and metallicity of stars are important in understanding the migration history of the Galactic disc because they encode the birth radii of stars \citep{Sellwood_Binney2002,Minchev2018, Lu2024, Ratcliffe2024}. \citet{Frankel2020} quantified the strength of radial migration and heating in the Galactic disc and showed that radial migration in the Galactic disc is significantly more efficient than radial heating.
Results from analytic studies and simulations have shown that radial migration could reshape the chemical properties of the Galactic disc, such as the azimuthal abundance distribution of the stars \citep{DiMatteo2013}, the $[\mathrm{\alpha/Fe}]$-bimodality of the disc \citep{Sharma2021_alphabimodality}, and the age-metallicity distribution \citep{Haywood2024}. Therefore, it is important to understand better the driving mechanisms of radial migration to unveil the evolution of the Galactic disc.

Theoretical analysis and N-body simulations have shown that transient spiral arms can drive efficient radial migration of stars \citep{Sellwood_Binney2002, Roskar2008, Roskar2012}, and many studies have emphasised the importance of evolving spiral arms and bar-spiral resonance overlap \citep[][and the reference therein]{Minchev2006_evolvingspiral, MinchevFamaey2010, Minchev2011_resonanceoverlap, Solway2012, Daniel2019}. Giant molecular clouds and satellite galaxies also cause orbital scattering \citep{SchonrichBinney2009, Quillen2009, Carr2022}. \citet{Hamilton2024} recently argued that using spiral arms to produce the required relative coldness of the migration process observed in the Galactic disc \citep{Frankel2020} is challenging, partly because resonance overlap also heats stars during transportation. In response, \citet{sellwood_Binney2025} provided evidence that the heating may be overestimated by \citet{Hamilton2024} in their idealized simulations. Another important potential source of radial migration is the slowing of the Galactic bar. During deceleration, the orbital resonances of the bar (e.g. corotation and outer Lindblad) move outwards through the disc, trapping stars and dragging them outwards \citep{Chiba2021, ChibaSchonrich_2021}. This mechanism has been demonstrated in N-body simulated galaxies as an efficient way to transport stars \citep{Halle2015, Halle2018, Khoperskov2020, Haywood2024}. Crucially, \citet{Khoperskov2020} showed that stars transported with the corotation resonance of the bar had near-circular orbits when they left the corotation resonance (see also \citealt{Chiba2021}) meaning resonant dragging has exactly the required properties of the radial migration mechanism.

A key parameter of the Galactic bar is its angular frequency, or pattern speed $\Omega$. Past observations have constrained the pattern speed of the Galactic bar to around $\sim34-41$~$\mathrm{km\,s^{-1}\,kpc^{-1}}$ \citep{Portail2017, Sanders2019, Bovy2019, Binney2020, Kawata_2021, ChibaSchonrich_2021}, and more recently, $\sim32-35$~$\mathrm{km\,s^{-1}\,kpc^{-1}}$ \citep{ClarkeGerhard_2022, Dillamore2024, Zhang2024_patternspeed}. This puts the corotation radius of the Galactic bar at $R\sim6.5-7.5$~kpc given that the rotation velocity of the disc is $230-240$~$\mathrm{km\,s^{-1}}$ \citep{McMillan2017}. However, in the past, the Galactic bar may not have been rotating with the pattern speed we measure today. It has been demonstrated analytically and computationally that the Galactic bar could decelerate due to dynamical friction against the dark matter halo \citep[][and reference therein]{DebattistaSellwood2000, Athanassoula2003, Weinberg2007}. \citet{Chiba2021} and \citet{ChibaSchonrich_2021} used the kinematics and chemistry of the solar neighbourhood stars to constrain the deceleration rate of the Milky Way bar as $\eta = -\dot{\Omega}/\Omega^2\sim0.0025-0.0045$. A slowing bar has been shown to be crucial in Milky Way modelling \citep{Chiba2021, Dillamore2024_GC, Li2024, Yuan2024}. In addition, the slowdown rate of the bar is an important tracer of the kinematics of the dark matter halo \citep{Athanassoula2003}, which opens a new window to constrain the nature of dark matter.


In this work, we compare the age-metallicity plane observed in the solar vicinity \citep{Xiang_2022} to a test particle simulation with a decelerating bar and investigate how bar-driven radial migration affects the age-metallicity plane in the solar vicinity. We describe the data and quality cuts that we adopted in Section~\ref{sec:data}. We present the setup of the test particle simulation in Section~\ref{sec:sim}. In Section~\ref{sec:results}, we compare the age-metallicity plane of the test particle simulation to observation in Section~\ref{subsec:AMR}; we discuss the implications of this radial migration mechanism to the disc $\mathrm{[\alpha/Fe]}$-bimodality in Section~\ref{subsec:alpha}; and we infer the pattern speed evolution history of the Milky Way bar in Section~\ref{subsec:slowdownrate}. In Section~\ref{sec:discussion}, we discuss other implications and other predictions that we expect to observe in the Milky Way under the bar-driven radial migration scenario.

\section{Data} \label{sec:data}

We use the stellar parameter and isochrone age measurements of $\sim247\,000$ subgiant stars derived from LAMOST DR7 spectra by \citet{Xiang2019} and \citet{Xiang_2022}. The typical age uncertainty is $\sim 7.5\%$ for stars younger than $10$~Gyr and $\sim10\%$ for older stars. To obtain stellar kinematics, we use the astrometric measurements from Gaia DR3~\citep{Gaia_DR3}, geometric distances derived in \citet{BJ_2021}, and the line-of-sight velocity measurements from LAMOST \citep{Deng_2012_LAMOST}. 
We calculate the orbital parameters (e.g. eccentricity and guiding radius) of stars using the Milky Way potential in \citet{McMillan2017} using AGAMA \citep{Vasiliev2019}. To ensure reliable orbital parameters, we select stars with small parallax uncertainties and small radial velocity uncertainties, i.e. $\varpi/\sigma_\varpi>5$ and $\sigma_{v_{\rm los}}<10$~km~s$^{-1}$.
We remove stars with metallicity error $\sigma_{\rm [Fe/H]}>0.1$ or age error $\sigma_{\tau}>1$~Gyr. To select stars belonging to the Galactic disc, we keep stars with orbital $\mathrm{eccentricity}<0.5$, vertical height $|z|<1$~kpc, $\mathrm{[Fe/H]}>-1$ and $\tau<11.5$~Gyr. These values are consistent with the metallicity and age measurements of the spin-up epoch of the Milky Way~\citep{Miglio2021, Belokurov2022, Rix2022, Conroy2022, Chandra2023, Zhang2023, Queiroz2023, Belokurov2024, Zhang2024, Liao2024}. There are $\sim128\,000$ stars left after these selections and $\sim97\%$ of them reside within $2.5$~kpc of the Sun. 

\section{Simulation} \label{sec:sim}

We run a test particle simulation with a realistic galactic potential. We gradually release particles into the simulation to investigate the response of stars of different ages to a decelerating bar. We summarise the key information here and present the details in Appendix~\ref{appendix:sim}.

We set up a time-evolving galactic potential using a similar approach to \citet{Dillamore2024}. To ensure the realism of the simulation, we adopt the potential of the inner Galaxy from~\citet[\citealt{Sormani2022}] {Portail2017}. 
We grow the galactic bar $4$~Gyr into the simulation, increasing the galactic bar strength following the prescription in \citet{Dehnen2000}. We let the bar reach its maximum strength at $t_{\rm bf}=5$~Gyr and slow the bar pattern speed after this. We set the initial pattern speed to $80\,\mathrm{km\,s^{-1}\,kpc^{-1}}$ and adopt a constant slowdown rate $\eta = -\dot{\Omega}/\Omega^2=0.003$. We also adjust the bar length according to its pattern speed to ensure $R_{\rm CR}/R_{\rm bar}$ is constant as the bar length should increase during deceleration \citep{Athanassoula1992}. At the final time of $11.5$~Gyr, the pattern speed is $\sim34\,\mathrm{km\,s^{-1}\,kpc^{-1}}$, similar to the measurements of the Milky Way bar \citep{ClarkeGerhard_2022, Zhang2024_patternspeed}, with corotation radius at $R_{\rm CR}\sim6.8$~kpc. The exact pattern speed evolution is described in Appendix~\ref{appendix:sim}.

We initialise the phase space of the test particles using the quasi-isothermal disc model \citep{Binney2010}. We linearly increase the scale length of the disc model from 1~kpc to 3.5~kpc during the simulation to mimic the inside-out disc formation. The details for the initial phase space distribution of the generated test particles are in Appendix~\ref{appendix:sim}. Every $100$~Myr, we sample $10\,000$ particles from the disc model and add them to the simulation. To simulate the dynamical heating, we also give random velocity kicks to the stars by convolving their three Cartesian velocity components with an isotropic Gaussian distribution, $\mathcal{N}(1,0.03)$, every $100$~Myr. The resulting age-velocity dispersion relation is roughly consistent with Milky Way observations \citep{Sharma2021_avr}. We assign a metallicity to each particle according to their birth radius and age using the method described in \citet{Lu2024} to better compare observation and simulation. To mimic the spatial coverage of the data, we apply a simple spatial cut to the output of the test particle simulation. We place the Sun at $R_\odot = 8.2$~kpc from the galactic center \citep{BHGerhard2016} and an angle $-25^\circ$ relative to the orientation of the galactic bar \citep{Bovy2019}. Particles beyond $2.5$~kpc from the Sun are removed. We use this subsample of the simulated test particles in the following sections unless otherwise stated.

\section{Results} \label{sec:results}

In this section, we analyse the results of the test particles and compare them to the observations in the solar vicinity. 

\subsection{Age-metallicity sequences in the solar vicinity}\label{subsec:AMR}

\begin{figure*}
    \centering
    \includegraphics[width=0.98\textwidth]{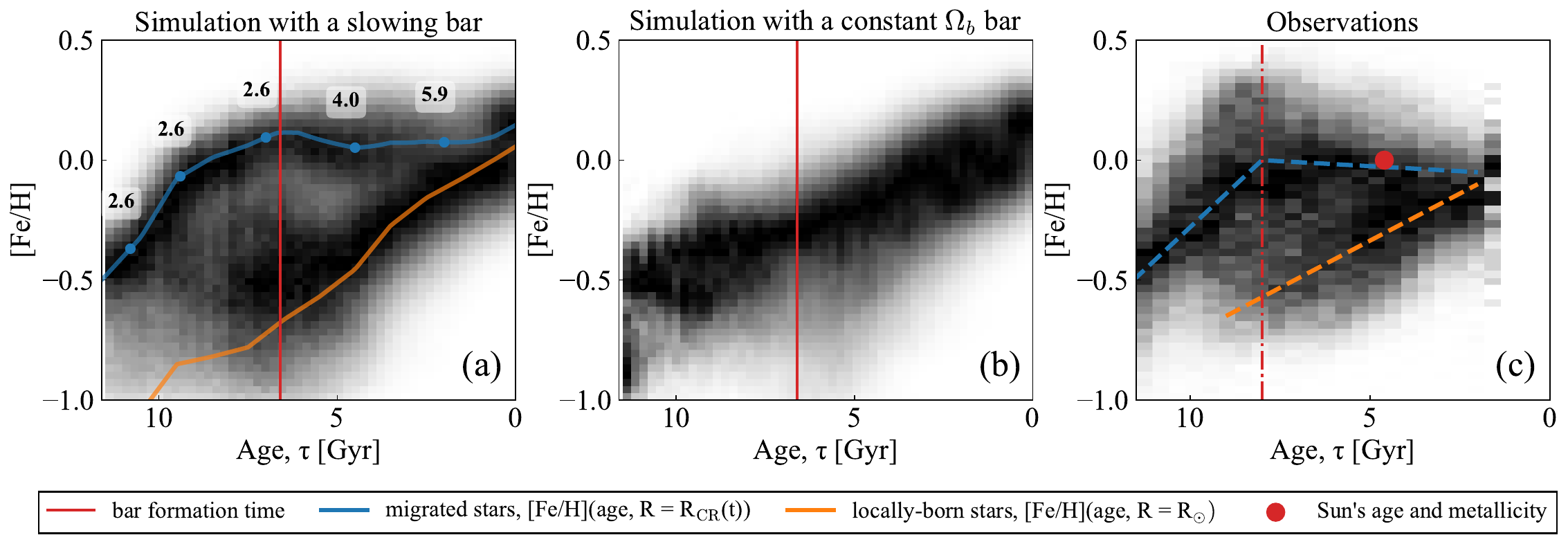}
    \caption{{\it Left, panel (a):} the age-metallicity plane of the test particle simulation with a slowing-down bar. The blue line shows the metallicity evolution for stars that formed at the corotation radius of their corresponding age, $\mathrm{[Fe/H]}(\tau,R_{\rm CR}(\tau))$, where the blue dots and annotation denote $R_{\rm CR}(t)$ at the respective times in units of kpc; the orange line shows the metallicity evolution of stars that formed around the solar circle, $\mathrm{[Fe/H]}(\tau,R_{\odot})$. The red line indicates the moment of bar formation, $t_\mathrm{bf}$. {\it Middle, panel (b):} the same as panel (a), except that the galactic bar is rotating with a constant pattern speed at all times.  {\it Right, panel (c):} the column-normalised age-metallicity distribution of the subgiant star sample we constructed based on \citet{Xiang_2022}, where the red circle indicates the Sun in this plane. The blue and orange dashed lines are used to denote the upper and lower age-metallicity sequences, and the red-dashed line labels the turning point of the upper sequence.}
    \label{fig:AMR_allstars}
\end{figure*}


Panel (a) of Fig.~\ref{fig:AMR_allstars} shows the column-normalised age-metallicity distribution of the simulated particles in the solar vicinity. Two clear sequences are apparent: an upper sequence that rises rapidly at early times and flattens for stars formed since the bar began decelerating (marked by the {\bf red}-solid line), and a lower sequence that exhibits a strong age-metallicity correlation all the way to young ages. We identify the two sequences with \emph{migrated} and \emph{local} stars respectively. The {\bf orange} solid line, $\mathrm{[Fe/H]}(t, R_\odot)$, represents the evolution of metallicity for particles that formed at the present solar radius, $R_\odot$, demonstrating that the \emph{local} sequence is composed of stars born around the solar circle.

The \emph{migrated} sequence is for particles that formed inside the solar circle and were dragged to the solar vicinity by the bar corotation resonance. The resonant dragging by the corotation of a decelerating bar increases the angular momentum of the trapped stars \citep{Chiba2021}, forcing them to migrate with the expansion of the corotation radius. This migration mechanism has also been identified in many N-body simulated galaxies \citep{Halle2015, Halle2018, Khoperskov2020}. For particles seeded before the bar formation, those that formed and reside around the initial corotation radius ($R_{\rm CR,0} = 2.6$~kpc in the simulation) are trapped at the moment of bar formation and are dragged later by the expansion of the corotation radius. This corresponds to the portion of the migrated sequence with a quick metallicity increase as a function of age. During bar deceleration, the corotation resonance expands and sweeps through the disc. The corotation resonance then traps particles born at larger radii, which have lower metallicity at the same age, causing the flattening in the age-metallicity sequence after bar formation. We use the blue solid line to denote the metallicity evolution for particles formed at the corotation radius at their birth time $R_{\rm CR}(t)$, i.e. $\mathrm{[Fe/H]}(t, R_{\rm CR}(t))$, where 
\begin{equation}
    R_{\rm CR}(t) =
    \begin{cases}
    R_{\rm CR,0}, & t < t_{\rm bf} \\
    V_{\rm c}/\Omega(t), & t > t_{\rm bf}.
    \end{cases}
    \label{eqn:RCRt}
\end{equation}
To summarise, the migrated sequence is composed of the age-metallicity relation at the initial corotation radius for the time before the bar formation and the segments of age-metallicity relations at the corresponding corotation radii at later times. The migrated sequence after the bar formation is flatter than the expected age-metallicity sequence at a fixed radius because stars at larger radii are dragged, but the extent of flattening depends on the details of the chemical evolution history (see Fig.~\ref{appendix:fig:chemcial_model} and Section~\ref{sec:boe_calc} for clarification).

As a comparison experiment, we run another test particle simulation that is the same as the previous simulation, except that the galactic bar is rotating with a constant pattern speed of $34$~$\mathrm{km\,s^{-1}\,kpc^{-1}}$. The resulting age-metallicity plane is shown in panel (b) of Fig.~\ref{fig:AMR_allstars}. 
A clear difference is that only one sequence is observed as the corotation resonance cannot drag particles from the inner galaxy to the outer disc, but instead only mixes stars around its corotation radius. Other non-axisymmetric perturbations (e.g., spiral arms) have to be involved in explaining the observation, which may lead to other inconsistencies \citep[e.g. see][]{Hamilton2024}.

\begin{figure*}
    \centering
    \includegraphics[width=0.95\textwidth]{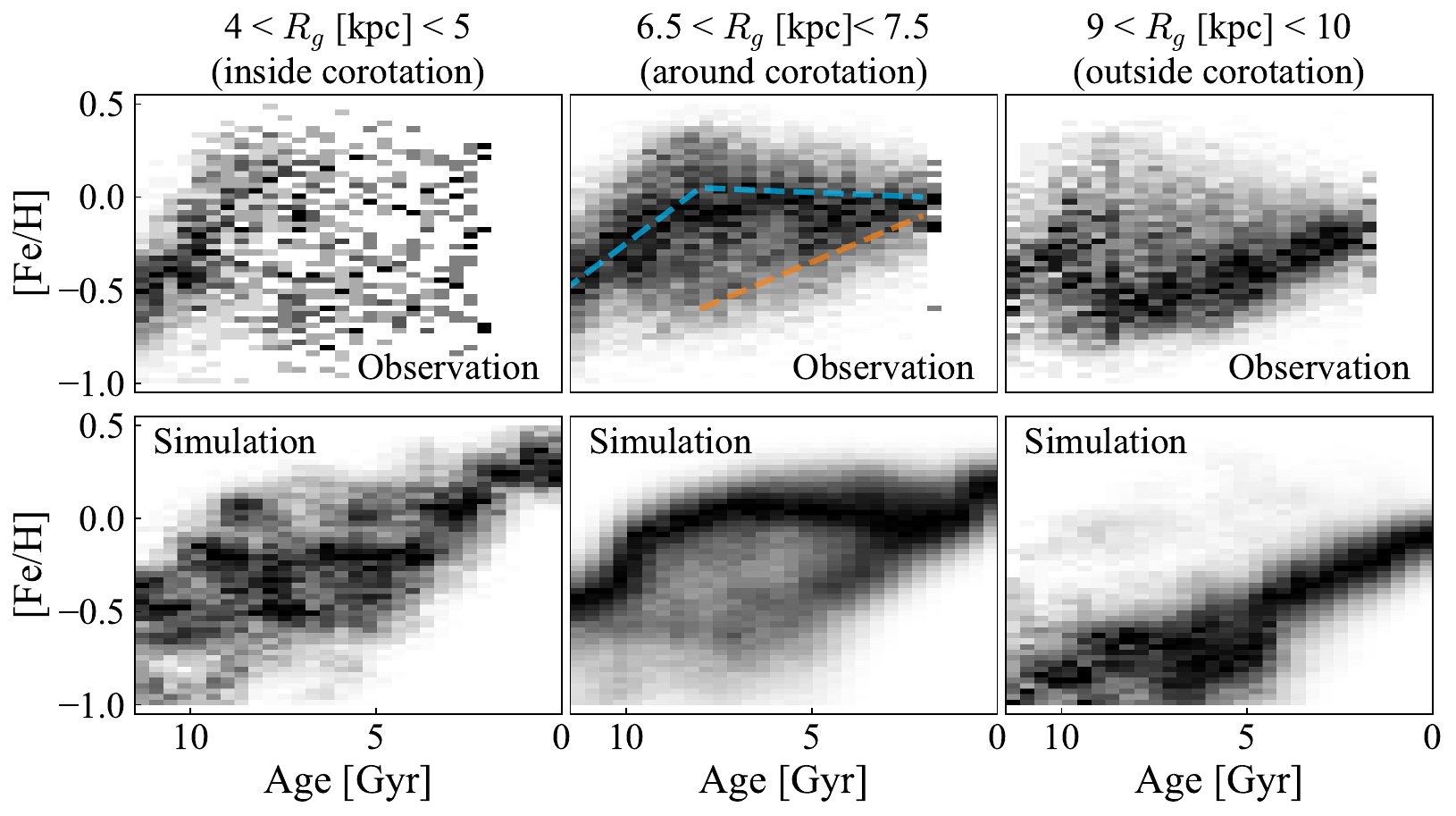}
    \caption{{\it Top panels:} the observed age-metallicity (column-normalised) distribution of the subgiant stars in various guiding radii bins. The blue and orange dashed lines in the middle panel denote the migrated and local age-metallicity sequence, which we will use in Fig.~\ref{fig:alphafe}. {\it Bottom panels:} the age-metallicity (column-normalised) distribution of the test particle simulation with a slowing down bar in different guiding radii bins.}
    \label{fig:AMR_guidingbins} 
\end{figure*}

In panel (c) of Fig.~\ref{fig:AMR_allstars}, we show the column-normalised age-metallicity distribution for the subgiant star sample described in Section~\ref{sec:data}.
As for the test particle simulation, there are two distinct sequences marked by the blue and orange dashed lines. The test particle simulation results suggest that the lower sequence is composed of stars that formed around $R\sim R_\odot$, and the upper sequence is composed of stars formed in the inner Galaxy that have migrated with the expansion of the corotation radius. \citet{Xiang_2022} also pointed out the migrated origin of the upper sequence. Under this scenario, the turning point in the upper sequence, indicated by the red dashed line, gives an estimate of the bar formation epoch of $\sim8$~Gyr ago, which is broadly consistent with other measurements of the bar formation time \citep{Sanders2024, Haywood2024}. We indicate the metallicity and age of the Sun with the red circle in panel (c). The Sun resides on the migrated sequence, suggesting that the Sun may have migrated to its current location with the corotation of the bar \citep{Baba2024}. A difference between panels (a) and (c) is that the upper envelope of the migrated sequence in the observations is decreasing whilst its counterpart in the test particle simulation is flat. This implies that either the deceleration rate is underestimated (see discussion in Section~\ref{sec:boe_calc}) or other non-axisymmetric perturbations neglected in the simulation have enhanced the radial migration.

\vspace{15pt}

\subsubsection{Age-metallicity sequences at different guiding radii}

We further examine the dependence of the sequences in the age-metallicity plane on the guiding radii, $R_g$, of stars. The upper (lower) panel of Fig.~\ref{fig:AMR_guidingbins} shows the column-normalised age-metallicity distribution of stars (test particles) in guiding radii bins of $4<R_g/\,\mathrm{kpc}<5$, $6.5<R_g/\,\mathrm{kpc}<7.5$, and $9<R_g/\,\mathrm{kpc}<10$, corresponding to stars (test particles) that are inside, around and outside the current corotation resonance, respectively (see the full plot of the age-metallicity distribution in all guiding radii bins in Fig.~\ref{appendix:fig:AMR_obs} and~\ref{appendix:fig:AMR_sim}). The distribution in the age-metallicity space strongly depends on the guiding radius in both the observation and simulation: 


\textemdash \ Inside the current corotation radius ($4<R_g/\,\mathrm{kpc}<5$, left column), we do not find a strong correlation in the age-metallicity plane, except for a sequence of increasing metallicity for stars older than $\sim8$~Gyr. This is the age-metallicity relation sequence for the high-$\alpha$ disc as discussed in \citet{Xiang_2022} (see fig.~2 therein). Fewer stars are observed at lower ages because of the imposed birth radii selection, which requires stars to have rather eccentric orbits to be born in the inner Galaxy and observed in the solar vicinity. As a result, the trends in the young stars are inconclusive. Similar shapes of the age-metallicity sequences have also been observed with the APOGEE data \citep{Lian2022_rm, Khoperskov2024}. The test particle simulation suggests that this guiding radii bin is dominated by stars born at that radius, but stars born inside and outside have non-negligible contributions as well. 

\textemdash \ For the observed stars with a guiding radius around the present-day corotation radius ($6.5<R_g/\,\mathrm{kpc}<7.5$, middle column), the upper age-metallicity sequence dominates, whereas the lower sequence also appears, but weakly. In the lower-middle panel, the age-metallicity plane of the test particles also shows the co-existence of these two sequences. 

\textemdash \ Outside the corotation radius ($9<R_g/\,\mathrm{kpc}<10$, right column), the lower sequence in both the observation and the simulation dominates the age-metallicity plane. A similar result is also presented in \citet{Lian2022_rm} using the APOGEE survey.

\begin{figure*}
    \centering
    \includegraphics[width=\textwidth]{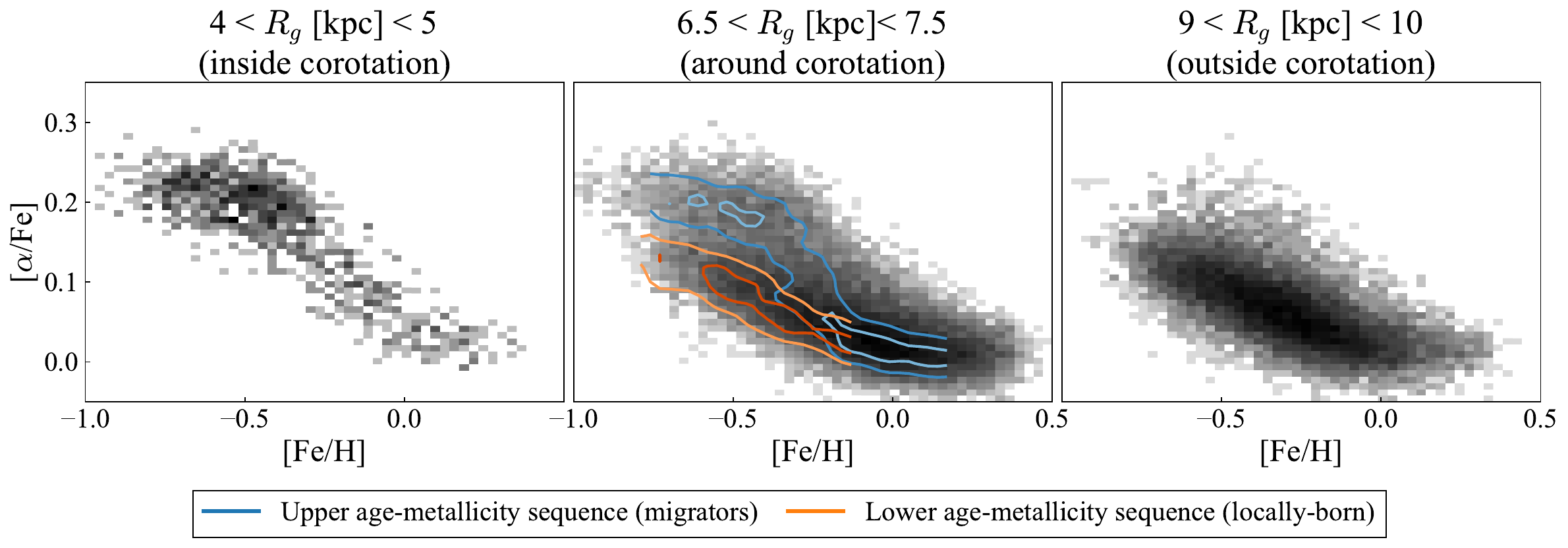}
    \caption{The (non-column-normalised) $\mathrm{[\alpha/Fe]-[Fe/H]}$ distribution of the observed stars in three guiding radii bins. In the {\it middle panel}, the blue(orange) contours show the column-normalised distribution of the stars in the upper(lower) age-metallicity sequence, which we selected in the top middle panel in Fig.~\ref{fig:AMR_guidingbins}.}
    \label{fig:alphafe} 
\end{figure*}

The dependence of the age-metallicity distribution on guiding radius sheds light on the origin of the two age-metallicity sequences. The observation that the upper sequence is the strongest around the corotation radius further supports our argument that the stars in the upper sequence are migrated due to the corotation resonance.
The guiding radii of stars trapped in corotation orbits oscillate around the corotation radii. The half peak-to-peak amplitude of this guiding radii oscillation in the potential model we used is $\sim 1-1.6$~kpc. Given that the present-day corotation radius is around $\sim6.5-7.5$~kpc \citep{Kawata_2021, ChibaSchonrich_2021, ClarkeGerhard_2022, Dillamore2023, Dillamore2024, Zhang2024_patternspeed, Cao2024}, we expect to see that the upper sequence exists only in the guiding radii bins between $5-9$~kpc, which is consistent with the observations (see Appendix~\ref{appendix:fullplot} for more plots).

\subsection{Implication for the disc $[\alpha/\mathrm{Fe}]$-bimodality}\label{subsec:alpha}

Radial migration driven by a decelerating bar also naturally explains the $[\alpha/\mathrm{Fe}]$-bimodality observed in the Galactic disc. 
We argued that the solar vicinity is occupied by stars with two different origins: the stars formed in the inner Galaxy that have experienced resonant dragging, and the stars formed around the solar circle. Due to the difference in the birth environment for these two groups of stars, the distribution of the $\alpha$-element abundance, $\mathrm{[\alpha/Fe]}$, at the same metallicity should be different. Under the inside-out disc formation scenario and the $[\alpha/{\rm Fe}]-$age relation of the Milky Way \citep{Miglio2021}, stars from the inner part of the Galactic disc are expected to be more $\alpha$-rich compared to the outer disc stars at the same metallicity.

An $[\alpha/\mathrm{Fe}]$-bimodality is reported in the LAMOST survey \citep{Xiang2019, Wang2022}. In Fig.~\ref{fig:alphafe}, we show the distribution of $\mathrm{[\alpha/Fe]-[Fe/H]}$ for stars in the same guiding radii as chosen in Fig.~\ref{fig:AMR_guidingbins}. We do not see an $[\alpha/\mathrm{Fe}]$-bimodality inside or outside the current corotation radius, as shown in the left and right panels. We have already discussed how the guiding radii bin of 4-5~kpc is impacted by the selection function which favours stars from the high-$\alpha$ thick disc. However, a similar result is also reported using the APOGEE survey, which should be less affected by this effect \citep{Hayden2015, Sharma2021_alphabimodality}. In the middle panel, which shows the $\mathrm{[\alpha/Fe]-[Fe/H]}$ distribution of stars around the corotation radius, the blue (orange) contour illustrates the column-normalised distribution of stars located around the upper (lower) age-metallicity sequence denoted by the blue (orange) dashed line in the upper middle panel of Fig.~\ref{fig:AMR_guidingbins}. The $\mathrm{[\alpha/Fe]-[Fe/H]}$ distribution naturally separates into two distinct sequences, the sequence of migrated stars (high-$\mathrm{[\alpha/Fe]}$), and the sequence of local stars (low-$\mathrm{[\alpha/Fe]}$). The high-$\mathrm{[\alpha/Fe]}$ migrated stars are born in the inner Galaxy ($R\lesssim4$~kpc, see the next section) before the bar starts to decelerate. With the slowing of the Galactic bar, the corotation resonance picks up stars at larger Galactocentric radii, which have lower $\mathrm{[\alpha/Fe]}$ values, causing a quick drop of $\mathrm{[\alpha/Fe]}$ for the migrated sequence at higher metallicity. The orange sequence is composed of the stars formed around the solar circle and hence has a lower $\mathrm{[\alpha/Fe]}$ value.


Our results supplement previous analyses that have shown that radial migration can reproduce the $[\alpha/\mathrm{Fe}]$-bimodality in the Galactic disc \citep{Sharma2021_alphabimodality, Chen2023}. The scale length growth of the disc has to be more finely tuned to match the observed radial dependence of the $\mathrm{[\alpha/Fe]-[Fe/H]}$ plane for radial migration mechanisms that do not favour the current corotation radius. Our analysis suggests decelerating bar-driven radial migration is the key mechanism responsible for the disc $[\alpha/\mathrm{Fe}]$-bimodality, which causes the migrated and local stars to co-exist around the present corotation radius, naturally. However, other effects in addition to radial migration could also cause $[\alpha/\mathrm{Fe}]$-bimodality in the Galactic disc, and our results cannot rule out their existence \citep[e.g.][]{Khoperskov2021, Agertz2021, Renaud2021}. 

\subsection{Inferred slowdown history of the Galactic bar} \label{subsec:slowdownrate}

We have developed a qualitative picture of bar-driven radial migration and now turn to a more quantitative estimate of the required bar properties to explain the migrated age-metallicity sequence dragged by the bar corotation resonance. We describe a motivational model that gives an initial estimate and builds intuition for the effect before giving a fuller calculation utilising a more realistic birth radii method.

\begin{figure*}
    \centering
    \includegraphics[width=\textwidth]{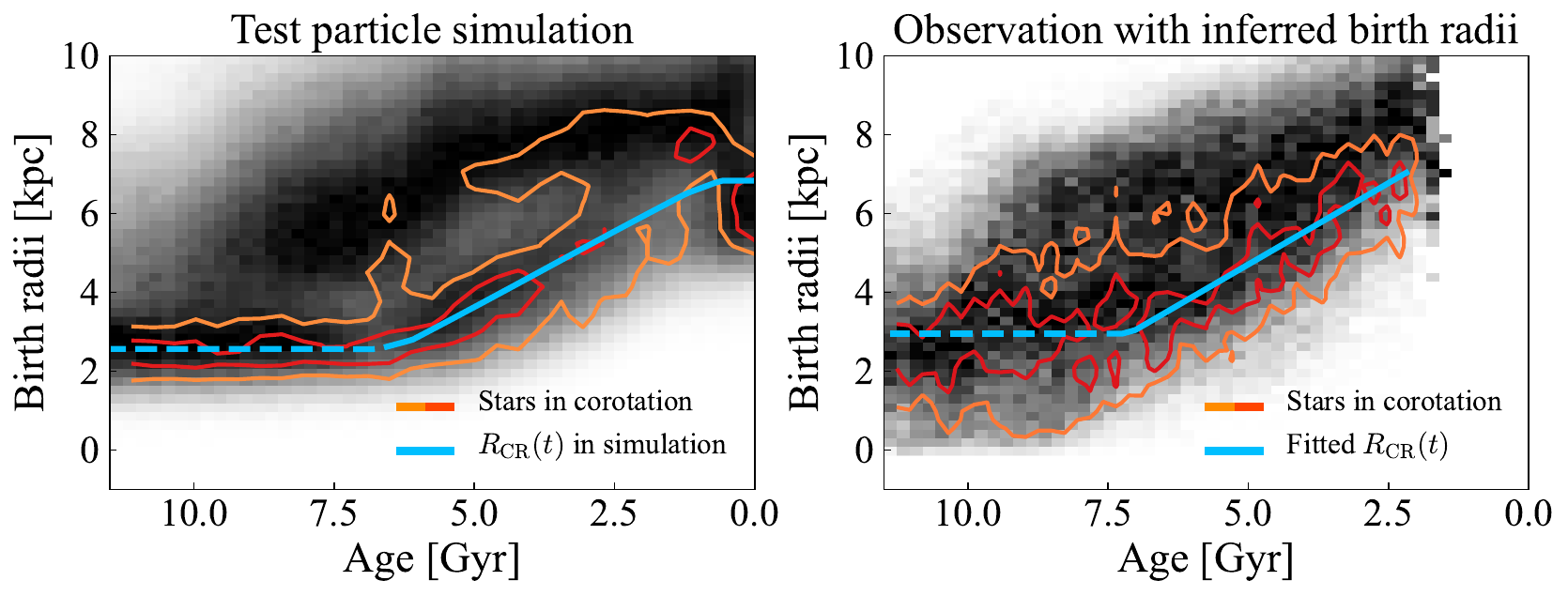}
    \caption{{\it Left panel:} the background is the column-normalised distribution of the test particles in the simulation in the age-birth radii($R_b$) plane, and the contour shows the same distribution of the selected particles that are corotating with the bar. The blue line indicates the temporal evolution of the corotation radius in the test particle simulation, in which the dashed part is for the time before the bar formation, and the solid line is after the bar formation. {\it Right panel:} the same to the left panel but for the observed stars with the birth radii inferred in \citet{Lu2024}, and the contour are for the selected corotation stars. The blue line is fitted for the sequences of the corotation stars, which we inferred the slowing-down history of the Galactic bar from it.}
    \label{fig:age_rb} 
\end{figure*}

\subsubsection{A motivational calculation}\label{sec:boe_calc} 
In a realistic galaxy model, the metallicities of stars depend on their birth time and radius, i.e. $[{\rm Fe/H}](t, R)$. Hence, the age-metallicity gradient of a sequence can be written as 
\begin{equation}
   \frac{D[\rm Fe/H]}{Dt} = \frac{\partial [\rm Fe/H]}{\partial R}\frac{\mathrm{d}R}{\mathrm{d}t} + \frac{\partial [\rm Fe/H]}{\partial t}.
   \label{eqn:agemetallicity_gradient}
\end{equation}
Under the scenario that the migrated (upper) age-metallicity sequence is a consequence of the dragging of the bar corotation resonance, we relate $\mathrm{d}R/\mathrm{d}t$ to the bar deceleration rate as 
\begin{equation}
    \frac{\mathrm{d}R_{\rm CR}}{\mathrm{d}t} = -\frac{\dot\Omega}{\Omega^2}{V_c} = \eta V_c.
\end{equation}
We then rearrange Eq.~(\ref{eqn:agemetallicity_gradient}) as 
\begin{equation}
    \eta = \frac{\partial {[\rm Fe/H]}/\partial t - D{[\rm Fe/H]}/Dt}{|\partial {[\rm Fe/H]}/\partial R|}\frac{1}{V_c},
    \label{eqn:eta_boe}
\end{equation}
We can estimate $\partial [\rm Fe/H]/\partial t$ from the slope of the local (lower) age-metallicity sequence in Fig.~\ref{fig:AMR_allstars}, which yields $\partial [\rm Fe/H]/\partial t\approx0.075\,\mathrm{dex\,Gyr^{-1}}$. We adopt $V_c\approx230~{\rm km\,s^{-1}}$ and $\partial [\rm Fe/H]/\partial R\approx-0.07\,\mathrm{dex/kpc}$~\citep{McMillan2017,Braganca2019}. As the age-metallicity gradient of the migrated sequence is almost flat, $D{[\rm Fe/H]}/Dt\approx0$, we calculate $\eta\approx0.0046$ from Eq.~(\ref{eqn:eta_boe}). From Eq.~(\ref{eqn:agemetallicity_gradient}), we can see that the flatness of the migrated age-metallicity sequence is a consequence of the balance between the bar deceleration rate and the metallicity enrichment rate. A decline or increase in the migrated age-metallicity sequence after bar formation could also be expected if the deceleration rate is faster or slower.

\subsubsection{Inference with the birth radii of stars}\label{sec:birth_radii}

We can decipher radial migration in the Galactic disc in greater detail with inferred birth radii of stars. 
With stellar metallicity and age measurements, \citet{Lu2024} improved on the method described in \citet{Minchev2018} and derived the empirical metallicity temporal evolution, $\mathrm{[Fe/H]}(R_b, \tau)$, with the same subgiant star sample that we used in this work \citep{Xiang_2022}. Using the metallicity evolution, they calculated the birth radii of these stars. The methodology is tested against various suites of cosmological simulations \citep[see][and some details in Appendix~\ref{appendix:Rb}]{Lu2022, Ratcliffe2024_tng50, Lu2024_lmc, Lu2024}. In our analysis, we adopt the birth radii inferred in \citet{Lu2024}. By projecting the age-metallicity plane to the age-$R_b$ plane, we are able to study the radial migration history of the Galactic disc in detail.

Corotation orbits are characterised by their special orbital shape, in which the corotation stars spend the majority of their orbital lifetime along one side of the Galactic bar. To select stars that are in corotation resonance with the Galactic bar, we integrate the orbits from their present-day phase space location for $\sim2$~Gyr, and we classify a star as trapped in the corotation resonance if its orbit resides on one side of the Galactic bar over $95\%$ of this time. The stars in the present-day corotation resonance are candidates for stars that have migrated with the decelerating bar. In the left panel of Fig.~\ref{fig:age_rb}, we show the column-normalised histogram of the age-$R_b$ plane for the test particle simulation in the background. The contours show the column-normalised distribution of the selected corotating stars. 
The blue line in the left panel indicates the temporal evolution of the corotation radius at different times in the test particle simulation, i.e. $R_{\rm CR}(t)$ in Eq.~(\ref{eqn:RCRt}). The blue line coincides with the distribution of trapped stars, demonstrating that we can use the birth radii of the corotating stars to infer the temporal evolution of the corotation radii in the Milky Way.

We repeat the analysis using the observed data and the birth radii inferred in \citet{Lu2024}. To select corotating stars, we integrate the orbits assuming that the current bar angle and pattern speed are $25^\circ$ and $34$~$\mathrm{km\,s^{-1}\,kpc^{-1}}$ \citep{Zhang2024_patternspeed}. In the right panel of Fig.~\ref{fig:age_rb}, we show the column-normalised distribution of all the stars in the background and that of the selected corotating stars by the contours. The contours for the corotating stars behave very similar to the test particle results, notably a plateau for the old stars and a linear increase for the young stars. We fit a function, $R_{\rm CR}(\tau)$, to the innermost contour of the selected corotating stars using the least-squares method, where
\begin{equation}
R_{\rm CR}(\tau) =
\begin{cases}
    R_{\rm CR,0}, & \tau > \tau_0, \\
    R_{\rm CR,0} + \eta V_c (\tau_0 - \tau), & \tau < \tau_0.
\end{cases}
\end{equation}
$R_{\rm CR,0}$ is the initial corotation radius, $\tau_0$ the age that the Galactic bar starts to decelerate, $\eta = -\dot{\Omega}/\Omega^2$ the deceleration rate, and $V_c$ the rotation curve which we fix the value to $230$~$\mathrm{km\,s^{-1}}$ \citep{McMillan2017}. The best-fit line is shown in blue in the right panel of Fig.~\ref{fig:age_rb}. The best-fit values are $R_{\rm CR,0}=2.9$~kpc, $\eta=0.0035$, and $\tau_0=7.2$~Gyr. The statistical uncertainties on these parameters are significantly smaller than the systematic uncertainties discussed later. The inferred slowing down rate is similar to the previous approximate calculation and is consistent with the measurement in \citet{Chiba2021}. The best-fit $\tau_0$ is consistent with previous measurements of the bar formation time \citep{Sanders2024, Haywood2024}. With the best-fit $R_{\rm CR,0}=2.9$~kpc, the initial pattern speed is then $\Omega_0\sim80$~$\mathrm{km\,s^{-1}\,kpc^{-1}}$. These parameters agree with the pattern speed evolution of Milky Way-type galaxies in the TNG50 cosmological simulation suite \citep{Semczuk2024}.


Since the history of the MW is mostly unknown, the birth radii inference in \citet{Lu2024} requires a few assumptions to obtain the metallicity evolution. Here, we experiment with adjusting the calibrated metallicity evolution model in \citet{Lu2024} to assess the systematic uncertainties in the deceleration rate inference. By shifting the present-day metallicity at the Galactic center ($\pm0.1\,\mathrm{dex}$) and the value of the steepest metallicity gradient ($\pm15\%$), we re-derive the birth radii of stars in our sample. We find that the general shape of the distribution of the corotating stars in the age-$R_b$ plane remains the same, but the exact values of the best-fit parameters vary. The parameters vary in the range of $\tau_0\sim6-8$~Gyr, and $\eta\sim0.0025-0.0040$, and $\Omega_0\sim60-100$~$\mathrm{km\,s^{-1}\,kpc^{-1}}$. The variation of the initial pattern speed is the highest as the value is sensitive to a small variation in $R_b$. These variations show the order of magnitude of the systematic uncertainty in our inference. We also vary the bar angle ($\pm 5^\circ$) and the present-day pattern speed ($\pm5\mathrm{km\,s^{-1}\,kpc^{-1}}$) needed to select the corotating stars, but these give variations in the best-fit parameters that are significantly smaller than the systematic uncertainties induced by the birth radii inference method. 

\section{Discussion} \label{sec:discussion}

\subsection{Predictions for the bar-driven radial migration scenario} \label{subsec:other_predictions}
In addition to the results shown above, there is more observational evidence to support the decelerating bar-driven radial migration as the main driver for the migration in the Milky Way disc. 

\begin{figure*}
    \centering
    \includegraphics[width=0.99\linewidth]{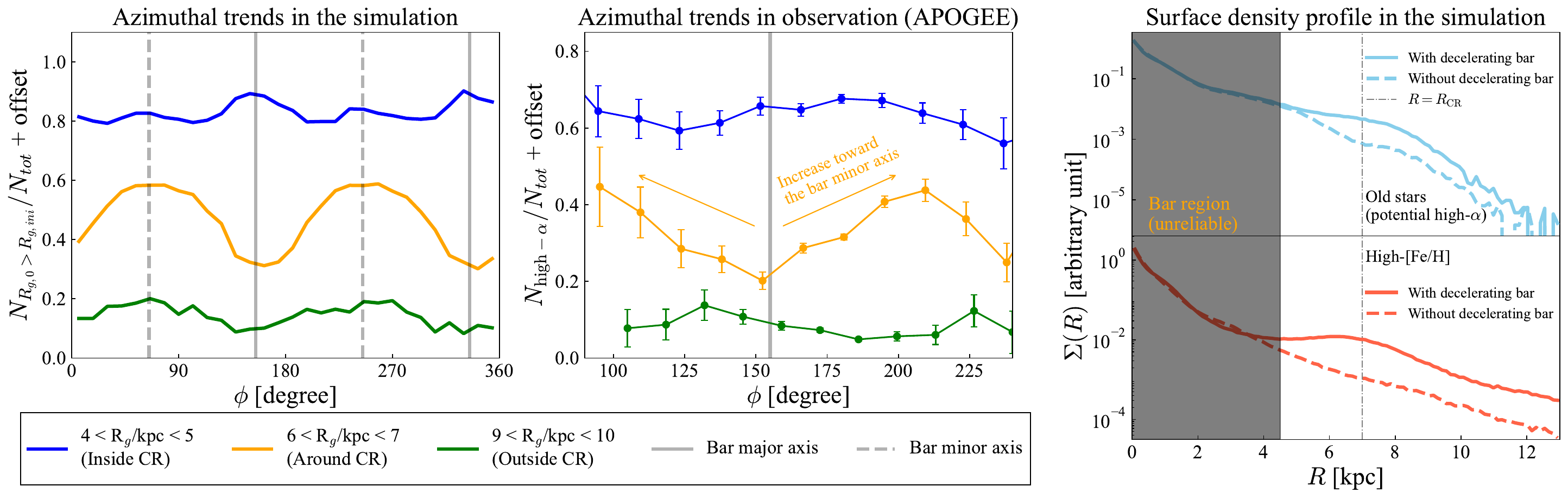}
    \caption{{\it Left:} the fraction of outward migrated stars as a function of azimuthal angle in different guiding radii bins. The verticle solid and dashed lines denote the bar major and minor axes. {\it Middle:} the observed fraction of high-$\alpha$ stars as a function of azimuthal angle. {\it Right:} the radial surface density profile of the old stars/high-[Fe/H] stars in the test particle simulation with a decelerating bar (solid lines) and with a constant pattern speed bar (dashed lines). The verticle dashed line represents the current corotation radius.}
    \label{fig:azimuthal_trends&density_profile}
\end{figure*}

{\it Azimuthal metallicity variation:} \citet{DiMatteo2013} showed that ongoing radial migration can cause azimuthal abundance variations in the galactic disc. A key difference in the bar-driven radial migration scenario from other migration mechanisms is the azimuthal distribution of the migrated stars. Due to the characteristic shape of the corotation orbits, corotating stars naturally clump along the bar minor axis around the corotation radius. Therefore, the stars that experienced corotation dragging and migrated from the inner Galaxy would form an overdensity along the bar minor axis as shown in the left panel of Fig.~\ref{fig:azimuthal_trends&density_profile}, causing the mean metallicity to peak along the bar minor axis. To demonstrate this, we use our decelerating bar test particle simulation without the solar neighbourhood cut to show the fraction of outward migrated stars as a function of azimuthal angle in different guiding radii bins, as shown in the left panel of Fig.~\ref{fig:azimuthal_trends&density_profile}. The orange line illustrates a clear azimuthal dependence in the spatial distribution of the migrated stars around the corotation resonance. Azimuthal metallicity variation has been observed in the Milky Way disc \citep{Poggio2022, Hawkins2023, Hackshaw2024}. Recently, \citet{Khoperskov2024} measured the metallicity distribution of the entire Galactic disc with orbital reconstruction. As shown in fig.~22 therein, the metallicity for stars around guiding radii of $7-8$~kpc varies sinusoidally in the azimuthal direction and peaks along the bar minor axes. Such an azimuthal trend is not seen in stars at other guiding radii. This result is in agreement with the prediction of our model. However, kinematic fragmentation \citep{Fragkoudi2017, Debattista2017, Zhang2024} could also cause a similar result, as kinematically cold (more metal-rich) stars are easier to be trapped. Therefore, more analysis is needed to differentiate these mechanisms. 

{\it Azimuthal variation in the high-$\alpha$ star fraction:} As we discussed in Section~\ref{subsec:alpha}, high-$\alpha$ stars observed in the solar vicinity are members of the migrated star sequence. Therefore, similarly to the azimuthal variation of metallicity, we might also expect an azimuthal variation for the high-$\alpha$ star fraction ($N_{\rm high-\alpha}$/$N_{\rm total}$), where more high-$\alpha$ stars are expected along the bar minor axes. As shown in the middle panel of Fig.~\ref{fig:azimuthal_trends&density_profile}, we examine this using a subsample of APOGEE DR17, following the quality cuts we addressed in detail in Appendix~\ref{appendix:apogee}. We indeed see an azimuthal variation in the fraction of high-$\alpha$ stars around the corotation radius, while the variation is much smaller outside the corotation radius. The fraction is smallest around the bar's major axis and increases towards the minor axis. The prediction is also in agreement with the results shown in \citet{Khoperskov2024}, who found that the high-$\alpha$ sequence contributes more in the direction of the minor axes than the major axes.

{\it Coldness of the Galactic disc:} \citet{Frankel2020} found that diffusion in the angular momentum of the Galactic disc stars is much greater than the scattering of radial actions. Recently, \citet{Hamilton2024} showed that it is difficult to transport stars with such coldness with transient spiral arms unless the lifetime and pitch angle for the spiral arms are fine-tuned \citep[but also see the counterevidence in][]{sellwood_Binney2025}. However, in the bar-driven radial migration scenario, the angular momentum of the corotating stars increases with the slowing of the bar, but the radial actions are invariant (see \citealt{Sellwood_Binney2002, Chiba2021} for theoretical explanations). Computationally, \citet{Khoperskov2020} showed that the orbits of stars transported with the corotation resonance can remain circular. So, the bar-driven radial migration scenario can alleviate the problem of the coldness of the migration, but more experiments or quantitative analysis are required to demonstrate which fraction of the slowing bar-driven migration is needed to explain the observation. 

{\it Density profile of the disc:} As argued, the Galactic high-$\alpha$ and high-$\mathrm{Fe}$ disc should also experience significant radial migration driven by the decelerating bar. 
Stars that experienced resonant dragging clump around the present-day corotation radius producing a maximum in the radial exponential surface density profile of the high-$\alpha$ and high-$\mathrm{Fe}$ (super-solar) disc leading to the broken-power law profiles as observed by \citep{Yu2021, Lian2022}. We demonstrate the role of the decelerating bar using our simulations in the right panel of Fig.~\ref{fig:azimuthal_trends&density_profile}. We show the surface density profile of the simulation with a decelerating bar (in solid lines) and with a constant pattern speed bar (in dashed lines) for an old population ($\rm age>8$~Gyr, potentially high-$\alpha$ stars in the Milky Way) and a high-[Fe/H] population ($\rm [Fe/H]>0.2$). The surface density profiles of both populations deviate from the initial exponential profile around the current corotation resonance in the presence of a decelerating bar. As other mechanisms may also lead to the broken power law density profile, more studies are required to build the unique correlation between the disc's density profiles and the bar-driven radial migration mechanism.

\subsection{Caveats of the test particle simulation} \label{subsec:caveat}

Our test particle simulations have a number of simplifications and limitations.

First, we kept the bar strength constant since the bar's formation. However, a realistic bar may experience a buckling instability during its evolution. When bar buckling occurs, the bar strength drops rapidly and many stars trapped by the corotation resonance may escape \citep{Khoperskov2020}. Hence, the efficiency of radial migration could have been overestimated in the test particle simulation. Nevertheless, the resonant drag of the galactic bar should still migrate the stars after the bar buckling, so it would not affect our argument. Similarly, the test particle simulation and our quantitative model in Section~\ref{sec:birth_radii} both assumed a constant slowdown, $\eta$, of the bar over its lifetime. Cosmological simulations \citep[e.g.][TNG50 as analysed by]{Semczuk2024} show a diversity of slow-down histories (and even periods of spin-up) so it is likely the Milky Way's bar history is more complex than our simple model. This simplifying assumption could then lead to an overestimate of the resonant trapping efficiency and hence the migration efficiency. 

Second, the initial conditions of stellar particles released to the test particles after bar formation are in disequilibrium with the potential in the bar region because the initial conditions are axisymmetric while the potential is barred. This leads to an unrealistic simulation result in the bar region, but stellar particles born outside the bar region are less affected by this issue as the non-axisymmetric component of the potential becomes weaker. As we mainly analyse the corotation resonance and stars around it, which is constantly outside the bar region in the simulation setup, this issue would not impact our conclusion much.

Third, we adopted a specific chemical model to assign the metallicity of the particles in the simulation. The detailed quantitative results presented in this work depend on the method in \citet{Lu2024}. However, the nature of our argument is that the decelerating bar-driven radial migration causes stars from very different birth radii to co-exist around the corotation radius, and the metallicity of the migrated sequence naturally flattens after the bar formation because the corotation resonance moves outward. Hence, switching to a different chemical model does not affect our argument as long as the radial metallicity gradient in the model is reasonable. Furthermore, our motivational calculation in Section~\ref{sec:boe_calc} and the observation of the break in the upper migrated age-metallicity sequence are direct empirical indicators which lead to very similar conclusions as the more detailed birth radii calculation.

\section{Conclusion} \label{sec:conclusion}

In this work, we analysed the age-metallicity distribution of subgiant stars in the solar vicinity and observed two distinct age-metallicity sequences (see panel (c) of Fig.~\ref{fig:AMR_allstars} for the observed age-metallicity plane). We ran a test particle simulation with a decelerating bar. The corotation resonance of the Galactic bar sweeps through the disc trapping particles. These trapped particles then experience resonant dragging and migrate with the expansion of the corotation radius. 
The simulation reproduces the two distinct sequences in age-metallicity: the upper age-metallicity sequence corresponds to the stars formed in the inner Galaxy that have migrated with the corotation resonance, while the lower sequence is composed of stars formed locally around the solar circle which have not undergone much radial migration. This explains the rapid metallicity increase at early times, the flattening at late times in the migrated sequence and the strong age-metallicity correlation in the local sequence.

We further showed the guiding radii dependence of the upper and lower age-metallicity sequences. The upper sequence is the dominant population around the present-day corotation radius, whereas the lower sequence dominates outside the corotation radius. We observed the upper sequence in the guiding radii bins of $\sim5.5-9$~kpc, which is consistent with the expected radial range of the current corotation resonance and the amplitude of the libration oscillation \citep{Chiba2021, ChibaSchonrich_2021}. We find the same guiding radii dependence in our test particle simulation, strengthening the conclusion that this sequence is composed of migrators dragged by the corotation resonance. 

We provided a series of other predictions if the slowing down of the Galactic bar is the main driver responsible for the radial migration in the Milky Way disc. We expect that: (1) the mean metallicity of stars is higher along the bar's minor axis around the corotation radius; (2) the fraction of high-$\alpha$ stars is the smallest along the bar's major axis and increases towards the minor axis around the corotation radius, which is demonstrated in Fig.~\ref{fig:azimuthal_trends&density_profile} using APOGEE survey; (3)
the migrated stars experienced much more radial migration than radial heating; 
(4) the radial density profile of the high-$\alpha$ and high-Fe stars should be consistent with a broken power law with the break close to the corotation radius. As discussed in Section~\ref{subsec:other_predictions}, many of these predictions are consistent with observations in the Milky Way.

We also shed light on the origin of the disc $\mathrm{[\alpha/Fe]}$-bimodality. Previous studies have discovered that radial migration could cause the $\mathrm{[\alpha/Fe]}$-bimodality of the disc \citep{SchonrichBinney2009,Minchev2013,Sharma2021_alphabimodality, Chen2023}. 
Here, we propose that migration driven by the decelerating bar could be the major mechanism, as the existence of the $\mathrm{[\alpha/Fe]}$-bimodality also shows a clear dependence on the guiding radii and is perhaps most prominent around the present-day corotation resonance of the bar. We show that the two distinct age-metallicity sequences occupy different regions in the $\mathrm{[\alpha/Fe]}-\mathrm{[Fe/H]}$ plane, resulting in the $\mathrm{[\alpha/Fe]}$-bimodality \citep{Khoperskov2024}. 

Finally, under the radial migration scenario, we use the ages and metallicities of stars currently trapped at corotation to infer the evolution history of the Galactic bar's pattern speed. We used the birth radii inferred in \citet{Lu2024} to track the temporal evolution of the corotation radius in the Milky Way's past. The results suggest that the Galactic bar formed and started to decelerate $\tau_0\sim6-8$~Gyr ago with an initial pattern speed of $\Omega_0\sim60-100\,\mathrm{km\,s^{-1}\,kpc^{-1}}$ and a constant slowdown rate $\eta\sim0.0025-0.0040$, consistent with previous constraints \citep{Chiba2021,ChibaSchonrich_2021,Sanders2024,Haywood2024}. 

\section*{Acknowledgments}
\begin{acknowledgments}
We thank Anke Ardern-Arentsen, Stephanie Monty, Francesca Fragkoudi, Payel Das, Eugene Vasiliev, and Chengdong Li for helpful discussions. We thank the anonymous referee for the careful review and helpful comments. 

HZ thanks the Science and Technology Facilities Council (STFC) for a PhD studentship. 
VB and NWE acknowledge support from the Leverhulme Research Project Grant RPG-2021-205: "The Faint Universe Made Visible with Machine Learning". 
JLS and AMD acknowledge support from the Royal Society (URF\textbackslash R1\textbackslash191555).
SGK acknowledges PhD funding from the Marshall Scholarship, supported by the UK government and Trinity College, Cambridge.
ZYL is supported by the National Natural Science Foundation of China under grant No. 12233001, by  the National Key R\&D Program of China under grant No. 2024YFA1611602, by a Shanghai Natural Science Research Grant (24ZR1491200), by the "111" project of the Ministry of Education under grant No. B20019, by the China Manned Space Project with No. CMS-CSST-2021-B03, and the sponsorship from Yangyang Development Fund.
\end{acknowledgments}

%

\vspace{5mm}


\software{Astropy \citep{astropy13, astropy18},
          SciPy \citep{Scipy}
          AGAMA \citep{Vasiliev2019}
          }



\appendix

\section{Details for the test-particle simulation}\label{appendix:sim}

In this appendix, we describe the details of the test particle simulation we used in this work.

\subsection{Time evolution of the galactic potential}
We follow the same approach as in \citet{Dillamore2024}, except that we use a different bar formation time. We adopt the Milky Way potential in \citet{Sormani2022}, which is an analytic approximation for the Made-to-Measure model of the inner Galaxy~\citep{Portail2017}. We use the non-axisymmetric part of this potential as a model of the galactic bar. We run the test particle simulation for $\sim11.5$~Gyr. The axisymmetric component of the potential is kept the same throughout the simulation so that the total mass is conserved. We let the bar component grow at $t_0\sim4$~Gyr since the start of the simulation and reach its maximum strength at $t_1\sim5$~Gyr, during which the pattern speed of the growing bar is constant. The smooth growth of the bar strength follows the same prescription as in \citet{Dehnen2000}. 

We set the initial pattern speed of the galactic bar as $\Omega_i=80\,\mathrm{km\,s^{-1}\,kpc^{-1}}$. The galactic bar starts to decelerate after the galactic bar reaches its full strength. We slow the galactic bar using a constant slowdown rate $\eta=0.003$ ($\eta = -\dot{\Omega}/\Omega^2$), which is roughly consistent with the value obtained for the Milky Way bar in \citet{Chiba2021}. The exact formula for the galactic bar pattern speed evolution is \citep{Chiba2021, Dillamore2024}:
\begin{align*}
    &\Omega(t) =
\begin{cases}
    \Omega_i &, t_0 < t < t_1 \\
    \Omega_i\left[1 + \frac{1}{2}\eta\Omega_i(t-t_1)^2/(t_2-t_1)\right]^{-1} &, t_1 < t < t_2 \\
    \Omega_2\left[1+\eta\Omega_2(t-t_2)\right]^{-1} &, t_2 < t < t_3 \\
    \Omega_4\left[1 + \frac{1}{2}\eta\Omega_4(t-t_4)^2/(t_3-t_4)\right]^{-1} &, t_3 < t < t_4 \\
    \Omega_4 &, t > t_4 \\
\end{cases}\\
    &\Omega_2 = \Omega_i\left[1 + \frac{1}{2}\eta\Omega_i(t_2-t_1)\right]^{-1} \\
    &t_3 = \frac{2}{\eta\Omega_4}-\frac{2}{\eta\Omega_2}+2t_2 - t_4, \\
\end{align*}
where the initial pattern speed is $\Omega_i=80$~km~s$^{-1}$~kpc$^{-1}$, final pattern speed $\Omega_4=34$~km~s$^{-1}$~kpc$^{-1}$, $t_0\sim4$~Gyr, $t_1\sim5$~Gyr, $t_2\sim5.5$~Gyr, and $t_4\sim11.2$~Gyr.
We also adjust the length of the bar scaled to its corotation radius so that the length of the bar matches its original value when the pattern speed is $39\,\mathrm{km\,s^{-1}\,kpc^{-1}}$ (the pattern speed value of the \citet{Portail2017} model). At the end of the simulation, the pattern speed is $34\,\mathrm{km\,s^{-1}\,kpc^{-1}}$.

\subsection{Initial condition of the test particles}
The initial conditions of the stellar particles are generated from a quasi-isothermal disc model $f(\bs J)$ \citep{Binney2010}:
\begin{align*}
&f(\bs{J}) = \frac{\tilde\Sigma\,\Omega_c}{2\pi^2\,\kappa^2} \times
\frac{\kappa}{\tilde\sigma_r^2} \exp\left(-\frac{\kappa\,J_r}{\tilde\sigma_r^2}\right) \times
\frac{\nu}   {\tilde\sigma_z^2} \exp\left(-\frac{\nu\,   J_z}{\tilde\sigma_z^2}\right) \times B(J_\phi),\nonumber\\
&B(J_\phi)=\left\{ \begin{array}{ll}  1 & \mbox{if }J_\phi\ge 0, \\
\exp\left( \frac{2\Omega_c\,J_\phi}{\tilde\sigma_r^2} \right) & \mbox{if }J_\phi<0, \end{array} \right.,\nonumber\\
&\tilde\Sigma(R_\mathrm{c})  \equiv \Sigma_0 \exp( -R_\mathrm{c} / R_\mathrm{disc} ) ,\nonumber\\
&\tilde\sigma_r^2(R_\mathrm{c}) \equiv \sigma_{r,0}^2 \exp( -2(R_\mathrm{c}-R_0) / R_{\sigma,r} ),\nonumber\\
&
\tilde\sigma_z^2(R_c) \equiv \sigma_{z,0}^2 \exp( -2(R_c-R_0) / R_{\sigma,z} ),
\end{align*}
where $J_{R}$, $J_{\phi}$ and $J_{z}$ are the radial, azimuthal and verticle actions, $\kappa$, $\nu$ the radial and verticle epicyclic frequencies, $\Omega_c$ the circular angular frequency. The adopted parameters are $R_0=8$~kpc, $R_{\sigma,R} = R_{\sigma,z} = 5$~kpc, and $\sigma_{r,0}=\sigma_{z,0}=10$~$\mathrm{km\,s^{-1}}$
We increase the scale length, $R_{\rm disc}$, of the disc model linearly from 1~kpc at the beginning of the simulation to 3.5~kpc at the end to mimic the inside-out disc formation. We generate stars using the sampling routine in \textsc{Agama} \citep{Vasiliev2019}. We release $10\,000$ particles into the test-particle simulation for every $\sim100$~Myr. 

\subsection{Dynamical heating}
In addition to the heating induced by the growing bar, we give a random velocity kick to the particles in the simulation by convolving their three Cartesian velocities with an isotropic Gaussian distribution $\mathcal{N}(1,0.03)$ every $\sim100$~Myr. This simulates the additional contribution of other scattering processes e.g. spirals, satellites etc. The resulting radial velocity dispersion for stars at $R=8$~kpc with age of $10$~Gyr in the test particle simulation is $\sim60$~$\mathrm{km\,s^{-1}}$, which is compatible with the observation in the Milky Way \citep{Sharma2021_avr}.

\subsection{Metallicity assignment}
With the birth radius and age of a star, we assign metallicities to the particles in the simulation using the metallicity enrichment derived in \citet{Lu2024}, $\mathrm{[Fe/H]}(R_b, \tau)$ where $R_b$ is the birth radius of the particle. The method is calibrated for the LAMOST subgiant sample in \citet{Xiang_2022}. the method assumes a monotonic increase in stellar metallicity at the Galactic centre and the radial metallicity gradient is constant throughout the time since the disc formation. In their method, the current metallicity gradient is -0.07~$\mathrm{dex\,kpc^{-1}}$, taken from \citep{Braganca2019}, and the highest metallicity is $0.6$~dex in the Galactic center. The steepest metallicity gradient is -0.15~$\mathrm{dex\,kpc^{-1}}$ at $8$~Gyr ago, similar to what is seen in cosmological simulations. A more detailed description of the method is provided in Appendix~\ref{appendix:Rb}. Using this method, metallicity could also serve as an indicator for the birth radii of the particle, for which more metal-rich stars are formed in the inner part of the galaxy at the same ages. The detailed metallicity growth at different radii is present in Fig.~\ref{appendix:fig:chemcial_model}. 

We directly use this observational calibrated method for the metallicity assignment in our test particle simulation for a better comparison between the observation and simulation, as we run the simulation from approximately the same age as the Milky Way disc formation. In Fig.~\ref{appendix:fig:chemcial_model}, we demonstrate that the (upper) migrated age-metallicity sequence follows the metal enrichment at the initial corotation radius before the bar formation and then crosses the age-metallicity relations at larger birth radii at a late time, causing its flattening.

\renewcommand{\thefigure}{A\arabic{figure}} 
\setcounter{figure}{0} 

\begin{figure}
    \centering
    \includegraphics[width=0.5\linewidth]{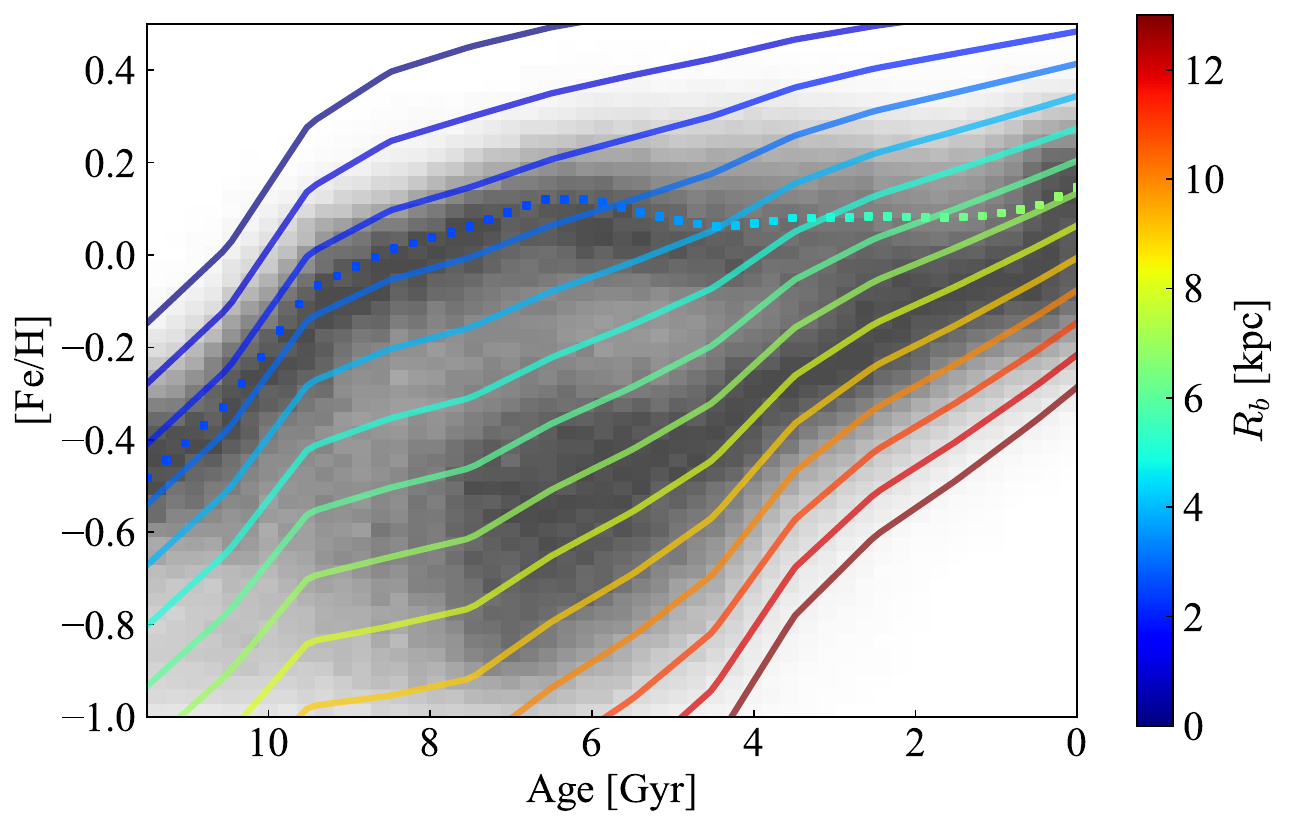}
    \caption{The model used for the metallicity assignment. Lines with different colour correspond to the metallicity evolution at different radii. The background shows the results of the test particle simulation with the decelerating bar. The dots denote the trend of $[\rm Fe/H](age, R_{CR})$.}
    \label{appendix:fig:chemcial_model}
\end{figure}

\section{Age-metallicity distribution in all guiding radii bins}\label{appendix:fullplot}

We present the age-metallicity distribution of stars and test particles in guiding radii bins of 4-5~kpc, 6.5-7.5~kpc, 9-10~kpc. Here, we show a more complete picture of the age-metallicity distribution with all guiding radii bins from 4-13~kpc with a width of 0.5~kpc. Fig.~\ref{appendix:fig:AMR_obs} shows that for the observations and Fig.~\ref{appendix:fig:AMR_sim} for the test particle simulation. The orange-dashed line in each panel of Fig.~\ref{appendix:fig:AMR_obs} is used to separate the upper and lower sequences, and the orange-dashed line in Fig.~\ref{appendix:fig:AMR_sim} is used to denote the expected occupation for the distribution of the particles transported by the corotation resonance (the same as the blue dashed line in Fig.~\ref{fig:AMR_allstars}).

The final corotation radius in the test particle simulation is $\sim7$~kpc similar to the expected value for the Milky Way ($\sim6.5-7.5$~kpc). The amplitude of the guiding radii libration oscillation is $\sim1-1.6$~kpc, and hence, the corotation stars are then expected to occupy the guiding radii range of $\sim5-9$~kpc. In both the observation and test particle simulation, the upper sequence exists ubiquitously among this range of guiding radii, and thus, agree with the expectation from the bar-driven radial migration scenario. 

\renewcommand{\thefigure}{B\arabic{figure}} 
\setcounter{figure}{0} 

\begin{figure*}
    \centering
    \includegraphics[width= \textwidth]{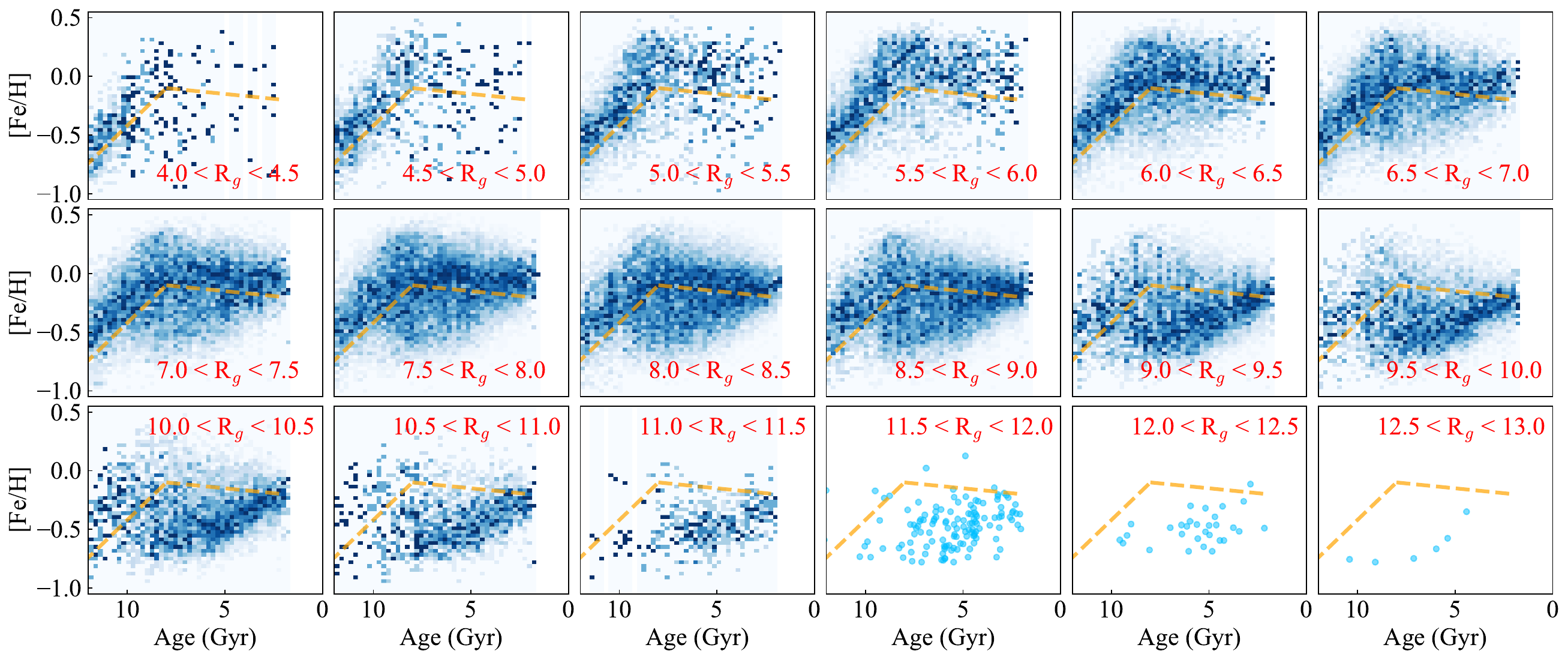}
    \caption{The age-metallicity distribution of the observed subgiant stars \citep{Xiang_2022} in different guiding radii bins, each with width of $0.5$~kpc. The orange dashed lines are used to denote the separation between the upper and lower sequence. }
    \label{appendix:fig:AMR_obs}
\end{figure*}

\begin{figure*}
    \centering
    \includegraphics[width= \textwidth]{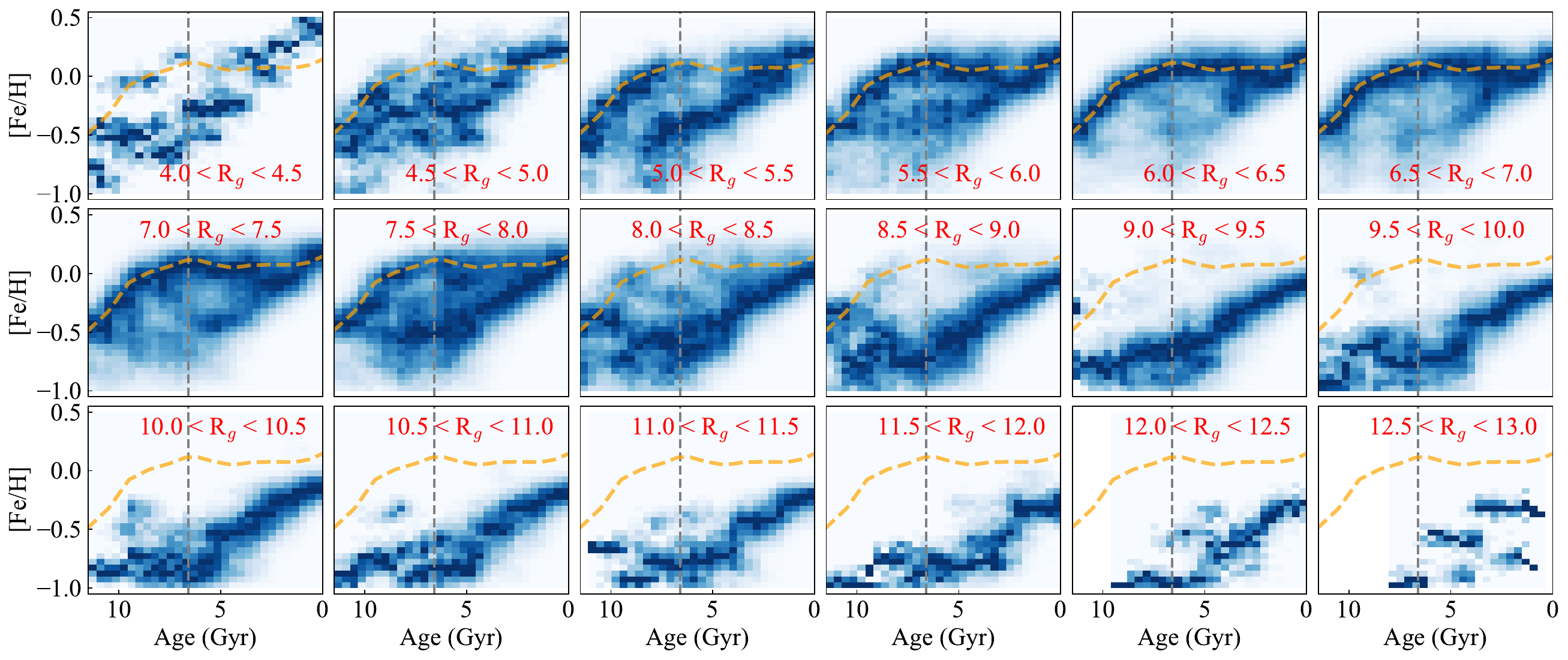}
    \caption{The age-metallicity distribution of the test particles (with a slowing down bar) in different guiding radii bins. The orange dashed lines are used to denote the sequence of particles migrated with the expansion of the corotation radius. We draw attention to the similarity between the observation and simulation around and outside the corotation.}
    \label{appendix:fig:AMR_sim}
\end{figure*}

\section{Birth radii calculation}\label{appendix:Rb}

The birth radii of stars are key information for decoding the radial migration in the Galactic disc. \citet{Lu2024} improved upon the method described in \citet{Minchev2018} to derive the birth radii of subgiant stars in \citet{Xiang_2022}. The method assumes that the radial metallicity gradient in the interstellar medium (ISM) is always linear after the stellar disc has started to form \citep[tested using cosmological simulations in][]{Lu2022}, which we then can write the stellar metallicity for a star with age, $\tau$, and birth radius $R_b$ as 
\begin{equation}
\mathrm{[Fe/H]}(R_b, \tau) = \mathrm{[Fe/H]}(R_b=0, \tau) + \nabla\mathrm{[Fe/H]}(\tau) \times R_b,
\label{appendix:eqn:feh_rbtau}
\end{equation}
where $\mathrm{[Fe/H]}(R_b=0, \tau)$ represent the central metallicity, and $\nabla\mathrm{[Fe/H]}(\tau)$ the metallicity gradient at the age. The central metallicity is estimated using the upper envelope of the observed age-metallicity plane at the corresponding age. \citet{Lu2024} postulated that the metallicity gradient at specific ages is linearly correlated with the stellar metallicity range observed at that age, $\Delta\mathrm{[Fe/H]}(\tau)$, based on the simulated galaxies in NIHAO-UHD \citep{Buck2020} and HESTIA \citep{Libeskind2020}. The linear correlation in written in the normalised metallicity range, $\Delta\widetilde{\mathrm{[Fe/H]}}$, which lies between 0 and 1, and 
\begin{equation}
    \nabla\mathrm{[Fe/H]}(\tau) = a \Delta\widetilde{\mathrm{[Fe/H]}} + b
\end{equation}
Recently, \citet{Ratcliffe2024} further corrected this correlation by adding the impact of the size of the star-forming region for the galaxy. The parameters in this linear correlation are determined using the boundary condition that the current metallicity gradient in the Galactic disc is $-0.07$~$\mathrm{dex\,kpc^{-1}}$ \citep{Braganca2019}, and the steepest metallicity gradient in the past. The latter is constrained by requiring that the youngest stars have close-to-zero migration and the disk forms inside-out on a reasonable timescale, which leads to a value $-0.15$~$\mathrm{dex\,kpc^{-1}}$, similar to what is found in cosmological simulations. With $\mathrm{[Fe/H]}(R_b, \tau)$, we can then rewrite Eq.~(\ref{appendix:eqn:feh_rbtau}) as 
\begin{equation}
    R_b = \frac{\mathrm{[Fe/H]_{\rm observed}} - \mathrm{[Fe/H]}(R_b=0, \tau)}{\nabla\mathrm{[Fe/H]}(\tau)} ,
\end{equation}

The assumptions and methodology are tested and verified against the NIHAO-UHD \citep{Buck2020} simulations in \citet{Lu2022}. It has also been tested on HESTIA \citep{Libeskind2020, Khoperskov2023}, TNG50 \citep{Nelson2019} for MW-like galaxies \citep{Ratcliffe2024_tng50}, and NIHAO galaxies \citep{Wang2015} down to the LMC mass \citep{Lu2024_lmc}.

\section{Azimuthal variation for high-$\alpha$ star fraction}\label{appendix:apogee}

We use the metallicity and $[\alpha/\rm M]$ measured in the APOGEE survey \citep{Abdurrouf2022} as it has a larger azimuth coverage than the LAMOST subgiant sample we used for the main results. Similarly to the data selection in Section~\ref{sec:data}, we keep stars with small distance uncertainty, small metallicity uncertainty, and small $[\alpha/\rm M]$ uncertainty, i.e. $\varpi/\sigma_\varpi>5$, $\sigma_{[{\rm M/H}]}<0.02$, and $[\alpha/\rm M]<0.05$. Stars with $[{\rm M/H}]>-1$ are removed for disc star selection. We only analyse stars with $0.5<|z|/{\rm kpc}<1.5$ to ensure that the $z$-distributions of the selected stars are roughly the same at each azimuthal angle.

We classify stars as a high-$\alpha$ star if $[\alpha/\rm M]>0.18$ for $[{\rm M/H}]<-0.7$ and $[\alpha/\rm M]>0.18-0.01([{\rm M/H}]+0.7)$ for $[{\rm M/H}]\geq-0.7$. We then take the ratio of the high-$\alpha$ to the total stars at different azimuthal angles in each of the guiding radii bins as shown in the middle panel of Fig.~\ref{fig:azimuthal_trends&density_profile}. There is an obvious increase of the high-$\alpha$ star fraction for the azimuthal angles away from the bar-major axis around the present-day guiding radii ($6<R_g/{\rm kpc}<7$), while the variation is much more insignificant inside and outside the corotation. This is in agreement with the expected consequence of the radial migration driven by the decelerating bar (see more discussion in Section~\ref{subsec:other_predictions}).

\bibliography{sample631}{}
\bibliographystyle{aasjournal}



\end{document}